\documentclass[intlimits,twoside,a4paper]{article}

\usepackage{amsmath,amssymb}
\usepackage{graphicx,subfigure}

\usepackage[T2A]{fontenc}
\usepackage[cp1251]{inputenc}

\usepackage[eqsecnum]{cmpj2}

\issue{2014}{17}{3}{33701}
\doinumber{10.5488/CMP.17.33701}

\newcommand{\figwidth}{0.45\textwidth}
\title[Self-organization of topological defects  for a  magnetic dots array]%
{Self-organization of topological defects \\ for a triangular-lattice
magnetic dots array \\ subject to a perpendicular magnetic field%
}
\author[R. Khymyn, V. Kireev, B. Ivanov]{ R.S.
Khymyn\refaddr{label2}, V.E. Kireev\refaddr{label1}, B.O. Ivanov\refaddr{label1}}
\addresses{
\addr{label2} Oakland University, 48309 Rochester Hills, MI, USA
\addr{label1} Institute of Magnetism, 03142 Kiev, Ukraine
}

\date{Received April 18, 2014, in final form May 15, 2014}
\authorcopyright{R. Khymyn, V. Kireev, B. Ivanov, 2014}

\begin{document}

\maketitle

\begin{abstract}
The regular array of magnetic particles (magnetic dots) of the form
of a two-dimensional triangular lattice in the presence of external
magnetic field demonstrates complicated magnetic structures. The
magnetic symmetry of the ground state for such a system is lower
than that for the underlying lattice. Long range dipole-dipole
interaction leads to a specific antiferromagnetic order in
small fields, whereas a set of linear topological defects appears
with the growth of the magnetic field. Self-organization of
such defects determines the magnetization process for a system
within a wide range of external magnetic fields.
\pacs 75.10.Hk, 75.50.Tt, 75.30.Kz
\end{abstract}

\maketitle

\section{Introduction and motivation}

Self-organization phenomena are usually associated with complex
systems far from equilibrium \cite{book1,book2,olem-book}. In that
case, the standard approach, depicting condensed matter as a gas of
weakly interacting quasiparticles (phonons, magnons, etc.) obtained
within the framework of linearized theory, may become inappropriate, and
the main role is played by soliton-type excitations
\cite{olem-book,KIK-book}. One can point out numerous examples of
such a behavior in very different physical systems. In particular, the
Berezinskii-Kosterlitz-Thouless phase transition in
quasi-two-dimensional magnets is driven by the appearance of a
finite density of free magnetic vortices
\cite{berez,kostThoul,kost74}. Plastic deformation of solids may
lead to the emergent complexity of a defect structure, resulting in a
hierarchy of super-defects such as dislocation and disclination
systems \cite{olemUFN92}. A fast quench across the phase transition
line may produce a finite density of frozen topological defects of
various types; such a behavior is known for a wide range of systems
from all fields of physics, either for classical \cite{olemUFN98} or
for quantum \cite{Dziarmaga} systems. Note the defect line scenario
of phase transitions for two-dimensional systems with discrete
symmetry breaking, see, for example, \cite{KorshunovUFN}. The common
feature of all the above examples is a considerable role of
stochastic factors, connected primarily to thermal fluctuations,
which leads to an irregular distribution of defects
\cite{olemUFN98,olemUFN01}.

Magnetic ordering is usually attributed to the exchange interaction of
atomic spins, leading to rather simple magnetically ordered states
\cite{SW}. Long-range magnetic dipole interaction usually produces
smooth non-uniformity (domain structures of different kinds) above
this simple exchange structure \cite{Hubert,BarYablUFN,BarIvJETP77}.
However, the theoretical investigation of the systems of magnetic
moments with pure dipolar interaction, the so-called \textit{dipolar
magnets}, shows that many physical properties, lacking in the
spin-exchanged systems, are present for those systems. We should first note
the presence of a non-unique ground state with nontrivial continuous
degeneracy for quite simple bipartite lattices, such as
three-dimensional cubic lattice, \cite{LutTisza,BelobGIgnat} and for
a two-dimensional square lattice, \cite{DipD2,DipD2',dipObzorUFN} as well as
specific phase transitions induced by an external magnetic field
\cite{BishGalkIv,GalkinIvMerk,GalkinIvSF}.

The models of dipolar magnets were originally discussed in regard to
real crystalline spin systems. A renewed interest to such models
has been caused by investigation of two-dimensional lattices of
sub-micron magnetic particles (the so-called magnetic dots), see
\cite{Skomski,Skomski2,IvFNT05} for a recent review. The ability to create
the dots with practically exactly equal sizes and precisely
controlled distance between them, leads to the long-range order
phenomena and can be treated as the creation of artificial crystals.

A direct exchange interaction between the dots is negligible, and
the dipolar interaction is the sole source of coupling between the dots.
For small enough dots of a size smaller than 100~nm, the
magnetization inside a dot is almost uniform, producing the total
magnetic moment $m_0 \gg\mu _\textrm{B} $, where $\mu _\textrm{B} $ is the Bohr
magneton, i.e., the typical value for an atomic magnetic moment. For
rather small magnetic dots of volume $10^3\div10^5$~nm$^3$, the value of
$m_0$ exceeds $10^4\mu _\textrm{B} $, and for dense arrays the characteristic
energy is higher than the energy of thermal motion at room
temperature \cite{BishGalkIv,PolitiPini02,GalkinIvMerk}.
Moreover, the Mermin-Wagner theorem is not valid for
two-dimensional magnets with a dipolar coupling of spins having
continuous degeneracy, and a true long range order can exist even
for a purely two-dimensional case at finite temperatures, either for
ferromagnets \cite{Maleev,Bruno} and antiferromagnets
\cite{IvTartPRL}. Thus, one can expect that thermal effects are
less important for magnetic dot arrays up to high temperatures
compared with the Curie temperature of the magnetic material.

Such systems represent dipolar magnets and fill their theoretical
investigation with a new physical content. Owing to the absence of
exchange, magnetic dot arrays constitute a promising material for
high-density magnetic storage media. For this purpose, the dense
arrays of small enough magnetic dots with magnetic moments
perpendicular to the array plane are optimal, see Refs.
\cite{chou94,Meier98,Ross}. Currently, the ordered arrays of magnetic
sub-micron elements have been discussed as materials for the so-called
\textit{magnonics}. In this new field in the applied physics of
magnetism, magnon modes with a discrete spectrum present for magnetic
nanoelements are used for processing the microwave signals
\cite{Magnonics}. Novel prospects are opened by the observation
of the excitation of collective spin oscillations by femtosecond
laser pulses \cite{SatohNatPh,kruglyak}. Magnon spectra for dot
arrays demonstrate a non-analytic behavior either for small
wavevectors,
\cite{GalkinIvZasp06,BondGIvZasp,Awad+gus,SukhostSpectr} or at some
symmetrical points within the Brillouin zone \cite{BondGIvZasp}.
Non-reciprocal effects for magnetic dot lattices
\cite{slavin,Fraerman} and non-trivial properties of the magnon
modes localized on the defects of dot arrays have been recently found
\cite{verba}.  The presence of phase transitions opens novel
opportunities for design of magnonic devices with the band structure
operated by external parameters, e.g., magnetic field, see for the
recent review \cite{KrawczGrObz}. Thus, magnetic dot arrays are
interesting as radically new objects for both the fundamental and
applied physics of magnetism.

If the magnetic moment of an individual magnetic particle is
perpendicular to the array plane ($xy$-plane), $\mathbf{m} = \pm
m_0\mathbf{e}_z$, the system can be described on the basis of the
Ising model. The energy of dipolar interaction of Ising moments
perpendicular to the system's plane is minimal for antiparallel
orientation of magnetic moments. Within the nearest-neighbors
approximation, such interactions lead to antiferromagnetic (AFM)
structures, e.g., simple chessboard AFM ordering is known for
a two-dimensional Ising square lattice with dipolar interaction
\cite{BishGalkIv,IvFNT05}. However, the close-packed triangular
lattices of the magnetic dots are also frequently used in
experiments. In particular, these lattices of cylindrical particles
considerably extended in the direction normal to the array plane are
naturally obtained when the array is prepared by controlled
self-organization \cite{TreugMass}. However, the triangular lattice
is not  bipartite. Such a lattice with AFM interaction of the
moments is a typical example of frustrated antiferromagnets
\cite{GekhtObz}. It is worth  mentioning here that the behavior of a
dot array with in-plane anisotropy is essentially different from
our case. In this case, the dipolar interaction is not clearly AFM;
and for a model of infinite unbounded array, the ferromagnetic state
is stable \cite{dipObzorUFN}, whereas for finite array, there appears a mesoscopic
non-uniform state of a form of either  domain wall
\cite{politiPiniStamps} or magnetic vortex
\cite{politiPiniStamps,DzianIv}.

For a nearest-neighbor Ising triangular lattice with AFM interaction,
thermodynamic properties are quite unusual \cite{GekhtObz}. It
is enough to mention that in this model there is no magnetic ordering
at any finite temperature $T \neq 0$; the ordering appears as a
result of accounting for the next-nearest-neighbor interactions only
\cite{NoPT,NoPT2,NoPT3}. This counterintuitive feature can be explained within
the concept of creation of linear topological defects with zero
energy \cite{Korshunov}. For the case of magnetic dipole
interaction, (common to what is observed for many next-nearest-neighbor
interaction models), our analysis shows the presence of AFM ordering
for small enough magnetic fields.

In the present work, a cascade of phases with different patterns of
dot magnetization has been found for a triangular lattice of
mesoscopic magnetic dots with perpendicular magnetization and in an
external magnetic field also perpendicular to the plane of the dot
lattice. The transition between these states is governed by
a novel mechanism involving the creation of an ordered system of linear
topological defects with non-zero magnetization. For those
transitions, thermal fluctuations are insignificant and one can
expect quite regular spatial distribution of such defects.

\section{Model description}

Consider a system of magnetic moments of magnetic particles
$\mathbf{m_n}$ placed in the sites of a triangular lattice
$\mathbf{n}$, parallel to the $xy$-plane,
\begin{equation}\label{D6z:W}
\mathbf{n}=ak\mathbf{e}_x +
\frac{al}{2}\left(\mathbf{e}_x+\sqrt{3}\mathbf{e}_y\right),
\end{equation}
where $a$ is a lattice constant, $k, l$ are integers, and
$\mathbf{e}_x$ and $\mathbf{e}_y$ are unit vectors parallel to $x$
and $y$ axes, respectively. We assume that any particle has
easy-axial (Ising-like) anisotropy of the form of $w_{{\bf{n}},a} =
{H_a}({m^2}_{{\bf{n}},x} +{m^2}_{{\bf{n}},y})/2{m_0}$, and that the
corresponding internal anisotropy field $H_a$  is substantially
higher than the characteristic field of the dot dipolar interaction,
$H_* = {m_0}/{a^3}$. Such systems of magnetic particles made
of soft ferromagnets, with the $H_a$  originating from the shape
anisotropy ${H_a}\sim 2\pi {M_\textrm{s}}$, where $M_\textrm{s}$ is the saturation
magnetization of the material, are used for magnetic memory
applications \cite{Ross}. The interaction field $H_*$  is
proportional to a small geometric factor, ${H_*} =
{M_\textrm{s}}({v_0}/{a^3})$, where $v_0$ is the dot volume. This interaction
field can be much less than $H_a$ even for dense enough lattices
\cite{BishGalkIv}; and it is not capable of deflecting the dot magnetic
moment from the $z$-axis.  Thus, we can assume that all magnetic
moments are ${\bf{m}}_{\bf{n}} = {m_0}{{\bf{e}}_z}{\sigma
_{\bf{n}}}$, where ${\sigma _{\bf{n}}} = \pm 1$. Then, the
Hamiltonian of this system of magnetic moments can be
written as
\begin{equation}\label{D6z:WW}
W = m_0^2 \sum_{\mathbf{n} \neq \mathbf{n}'} \frac{\sigma_\mathbf{n}
\sigma_{\mathbf{n}'}}{|\mathbf{n}-\mathbf{n}'|^{3}} - m_0 H
\sum_\mathbf{n} \sigma_\mathbf{n} \;,
\end{equation}
where an external magnetic field $\mathbf{H}=H \mathbf{e}_z$ is
applied perpendicularly to the plane of the array. The first term
describes dipolar interaction, with the summation performed over all
of the pairs of the lattice sites. Below, for the sake of simplicity,
we present the energy (per one magnetic particle) in the units
of $m_0^2 / a^3$ and we  use the dimensionless magnetic field,
$h=H/H_{*}$, where the characteristic value $H_{*} = m_0 / a^3$. The
present model applies directly to any triangular lattice of
identical dipoles that are restricted to the two directions of
normal orientation \cite{dipObzorUFN}. It is interesting that the
model formulated in this paper can be used to describe a system of
vortex state magnetic dots \cite{BishGalkIv} accounting for the
interaction of a magnetic moment of vortex cores. The direction of
the core moments is connected with the topological invariant, and
the vortices with upward or downward directions of the moments
survive until non-small values of magnetic field \cite{IvWysin02} are reached.

As has been mentioned above, two sources of
degeneracy are present in our system, and it is not \textit{a priori} obvious  which
structure will constitute the ground state. In this situation it is
natural to start with the numerical analysis of the problem.

\section{Ground states for infinite system: Numerical analysis}

To find the global minimum of the energy of a system, there was used
a Monte-Carlo (MC) approach combined with a simulated annealing (SA) method,
see the textbook \cite{MonteCarloBook} for details.
The standard MC method is based on the attempts to reverse the moment on the random site,
and the only reversals favorable to energy are allowed. By contrast,
for MC-SA, the probability of the reversal is non-zero even if the energy
grows after the reversal; otherwise, the system with a high probability
will be ``frozen'' in some local minimum.
The probability depends not only on the energy gain
$\Delta E$ but also on the  global parameter $T$ called the
\textit{temperature}. If the reversal is favorable in the energy, the
moment is always reversed, irrespective of $T$. But even if the
reversal is unfavorable, the non-zero probability of a reversal is
chosen as follows: flip-over takes place if $\Delta E < T |\log p|$,
where $T$ is the current value of temperature, $p$  is a random
value $0 < p \leqslant 1$. Here, the parameter temperature determines the
strategy of minimization and the meaning of the temperature is
the same as for annealing in metallurgy involving initial heating
and controlled cooling of a material, thereby avoiding
the formation of defects. $T$ changes according to the quantity of full steps
of MC-SA such that the initial temperature is high enough compared
with the interaction energy, and then the temperature decreases. For
a concrete analysis, we took the rhombus-shaped samples with
consecutively increasing periods up to $16 \times 16$ and used
periodic boundary conditions.

\subsection{Simplest ground states of the system: zero field and saturation }

\begin{figure}[!h]
\begin{center}
\includegraphics[width=\figwidth]{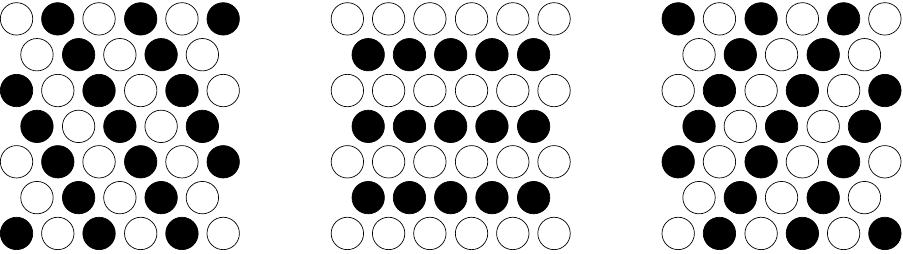}
\end{center}
\caption{Uniform antiferromagnetic states giving the energy minimum
at a small magnetic field. Here and below, in all figures, the open and
closed circles denote the particles with
the upward and downward moments, respectively.} %
\label{f:2-4states} %
\end{figure}

For a triangular Ising lattice in a several-neighbor approximation,
a simple AFM order with two sublattices can be implemented
\cite{SlotteHemmer84}. We found the same configuration  for a
long-range dipolar interaction in the absence of the field and for a
small enough magnetic field. For these states, the magnetic
elementary cell is rectangular having lower symmetry than for the
underlying triangular lattice, and this state possesses a much higher
discrete degeneracy than a simple chessboard structure for a
square lattice discussed before. Several AFM states can occur
in a system, which are different but fully equivalent by their
energies. Figure~\ref{f:2-4states} presents three of these states,
while the other three states are obtained from them by changing the
magnetic moment sign $\sigma_{\mathbf{n}}$ at all of the particles.
Note the deviation from the standard AFM case with the antiparallel
orientation of all the nearest neighbors, which is a typical
manifestation of frustration in a system.

\subsection{Monte-Carlo analysis for intermediate field values}

\begin{figure}[!t]
\begin{center}
\includegraphics*[width=\figwidth]{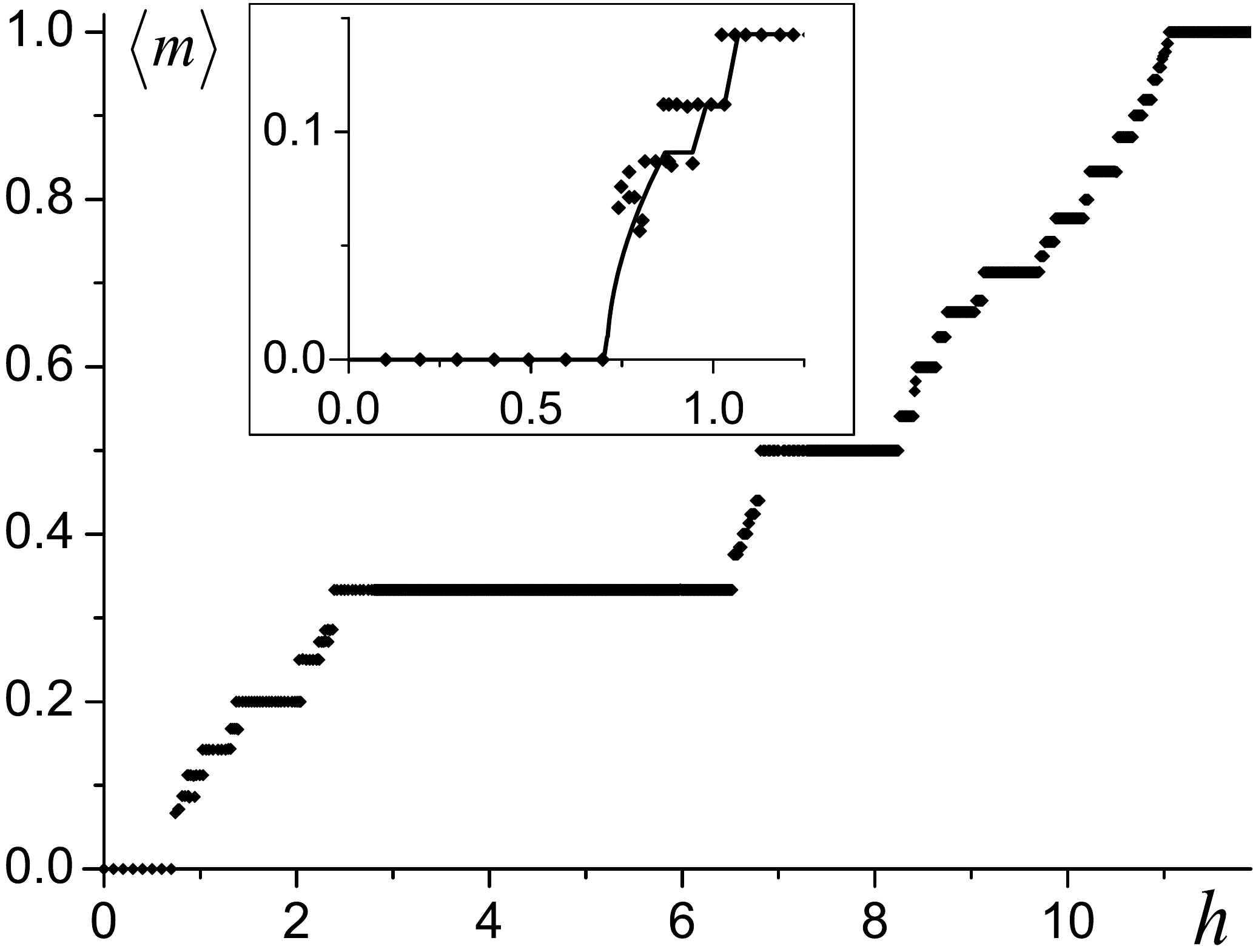}
\caption{The mean value of magnetization $\langle m_0 \rangle$ (in
units of $m_0$, per one dot) of the array as a function of magnetic
field (in units of $H_{*}=M_0/a^3$) found by Monte-Carlo
simulations. Magnetization function at low fields found by
an exhaustive search of the states of rhombus-shaped samples (full
line) together with  the Monte-Carlo data (symbols) are presented in
the insert. \label{f:D6z:magn_inf}}
\end{center}
\end{figure}

MC-SA analysis shows that for some small but finite values of the
magnetic field, at least up to $h = 0.7$, the mean value of the
magnetic moment $\langle m \rangle $ equals zero, which indicates a simple AFM structure.
For higher fields, numerous more complex
structures with $0 < \langle m \rangle < m_0$ occur in the
intermediate region between AFM state and saturated state. The mean
value of the magnetic moment (per one particle) $\langle m \rangle$
corresponding to these configurations, is present in figure~\ref{f:D6z:magn_inf}.

\begin{figure}[!b]
\subfigure[\ $h = 1.0$ \label{f:D6z:inf_p9}]
{
\includegraphics*[width=3cm]{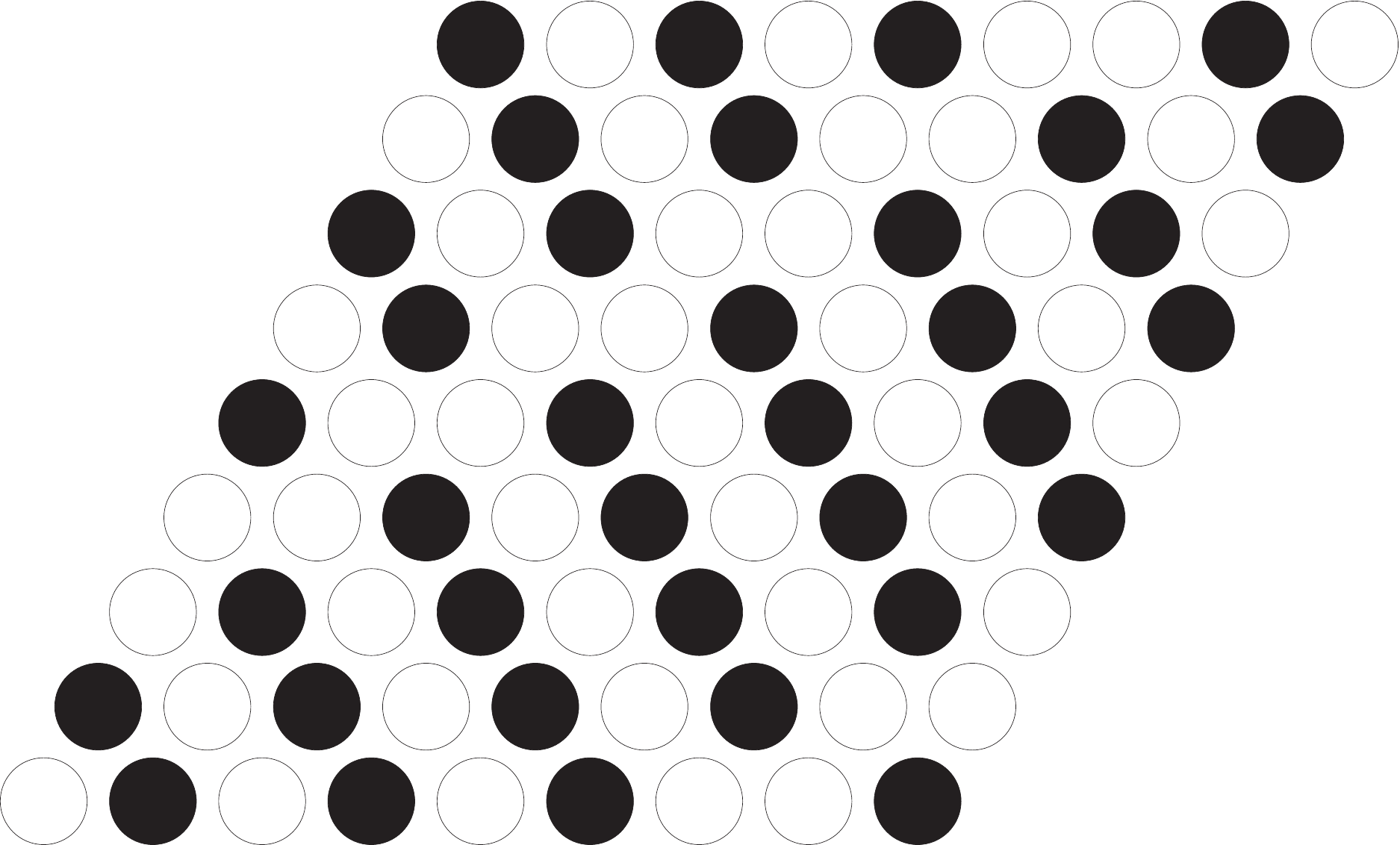}
}\hfill
\subfigure[\ $h = 1.1$ \label{f:D6z:inf_p7}]
{
\includegraphics*[width=3cm]{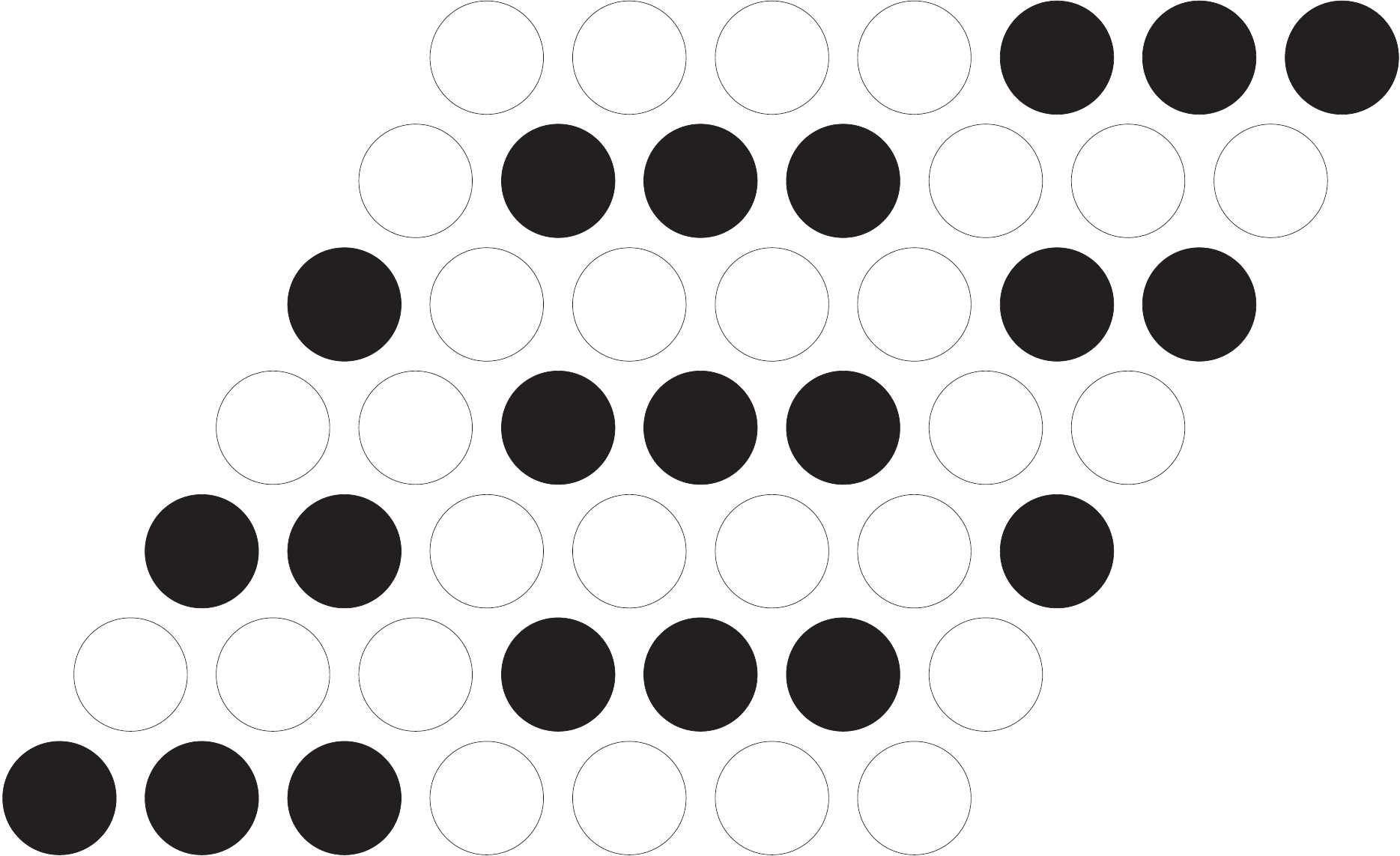}
}\hfill
\subfigure[\ $h = 1.5$ \label{f:D6z:inf_p5}]
{
\includegraphics*[width=3cm]{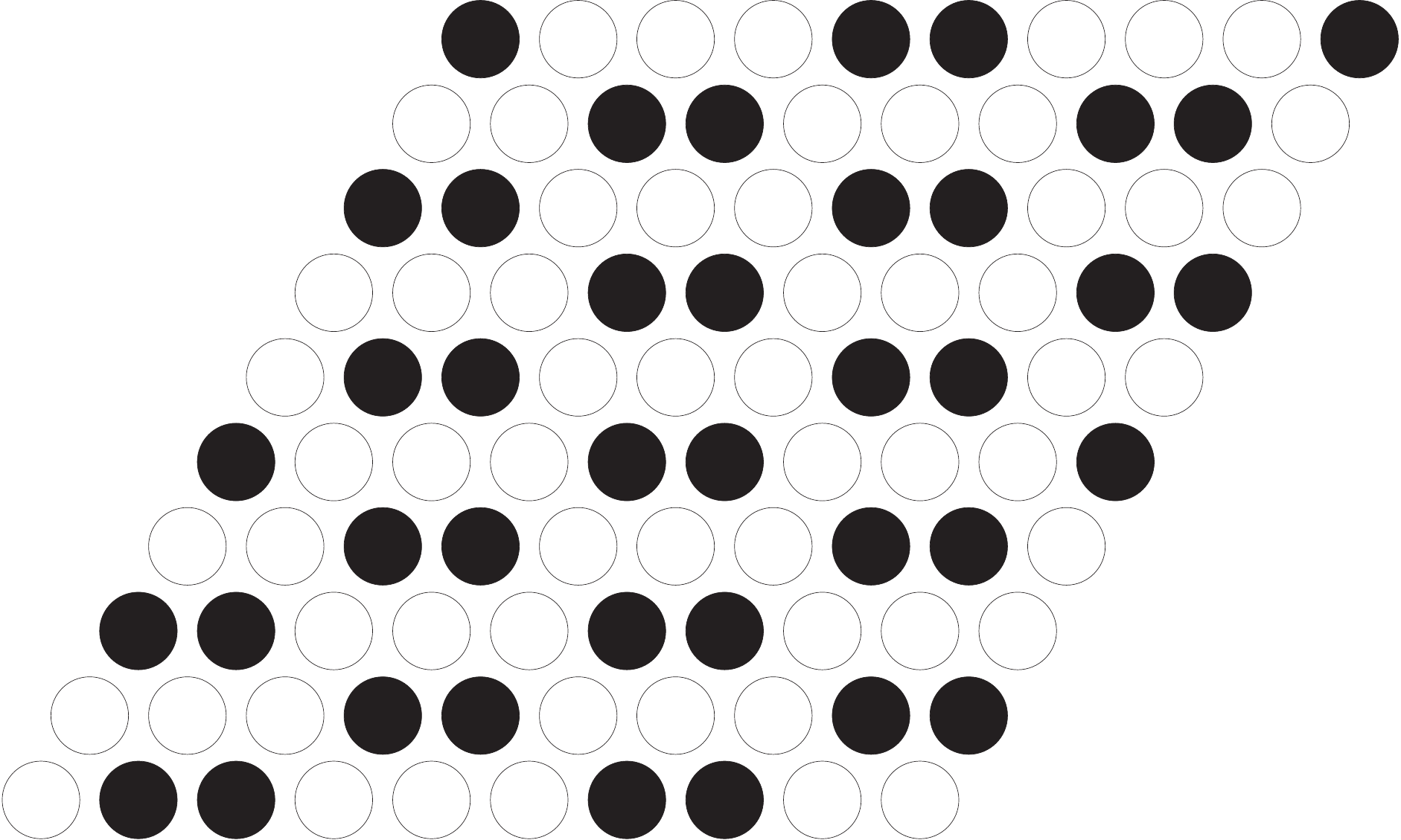}
}\hfill
\subfigure[\ $h = 2.1$ \label{f:D6z:inf_p8}]
{
\includegraphics*[width=3cm]{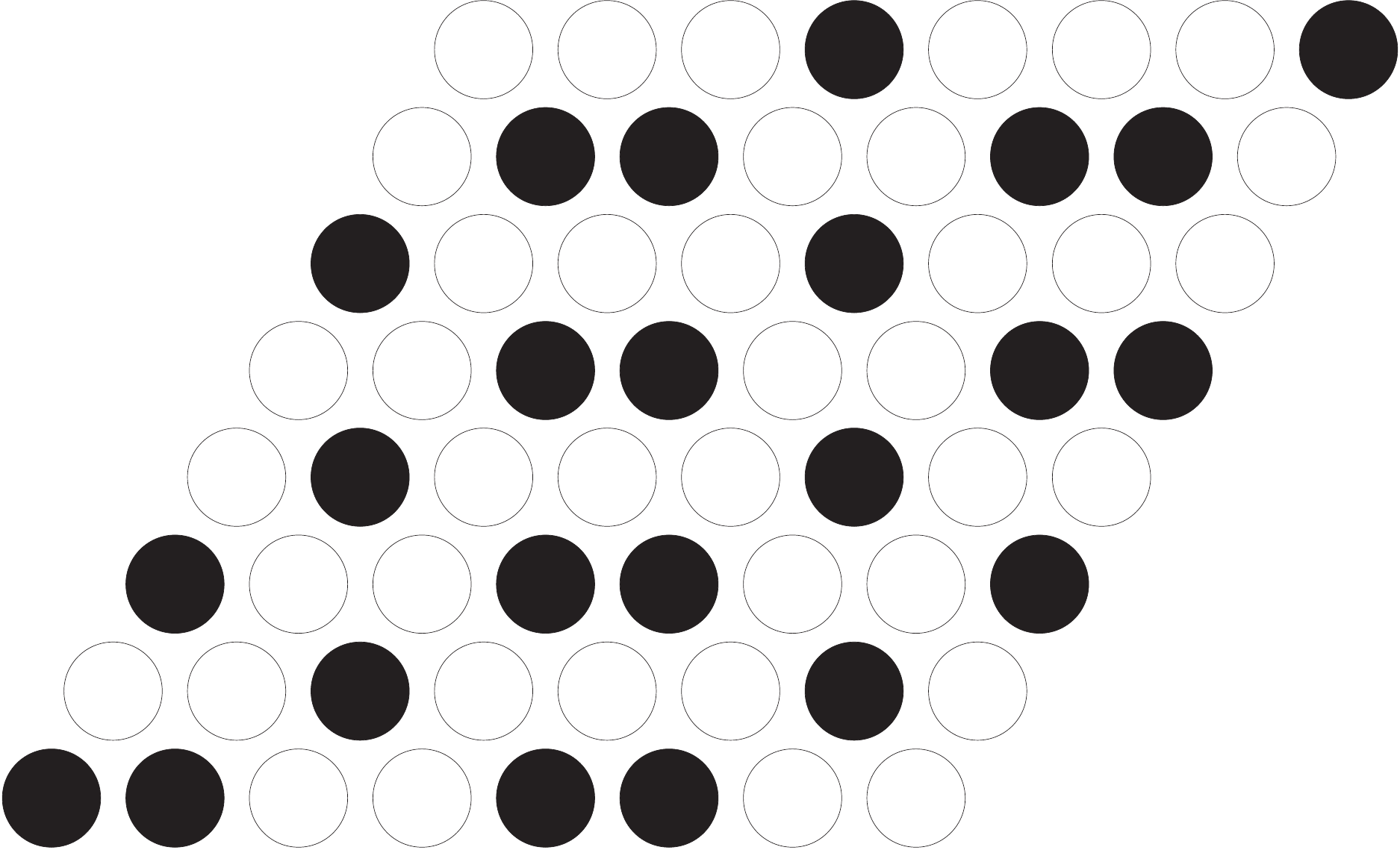}
}\hfill
\subfigure[\ $h = 2.3$ \label{f:D6z:inf_p11}]
{
\includegraphics*[width=3cm]{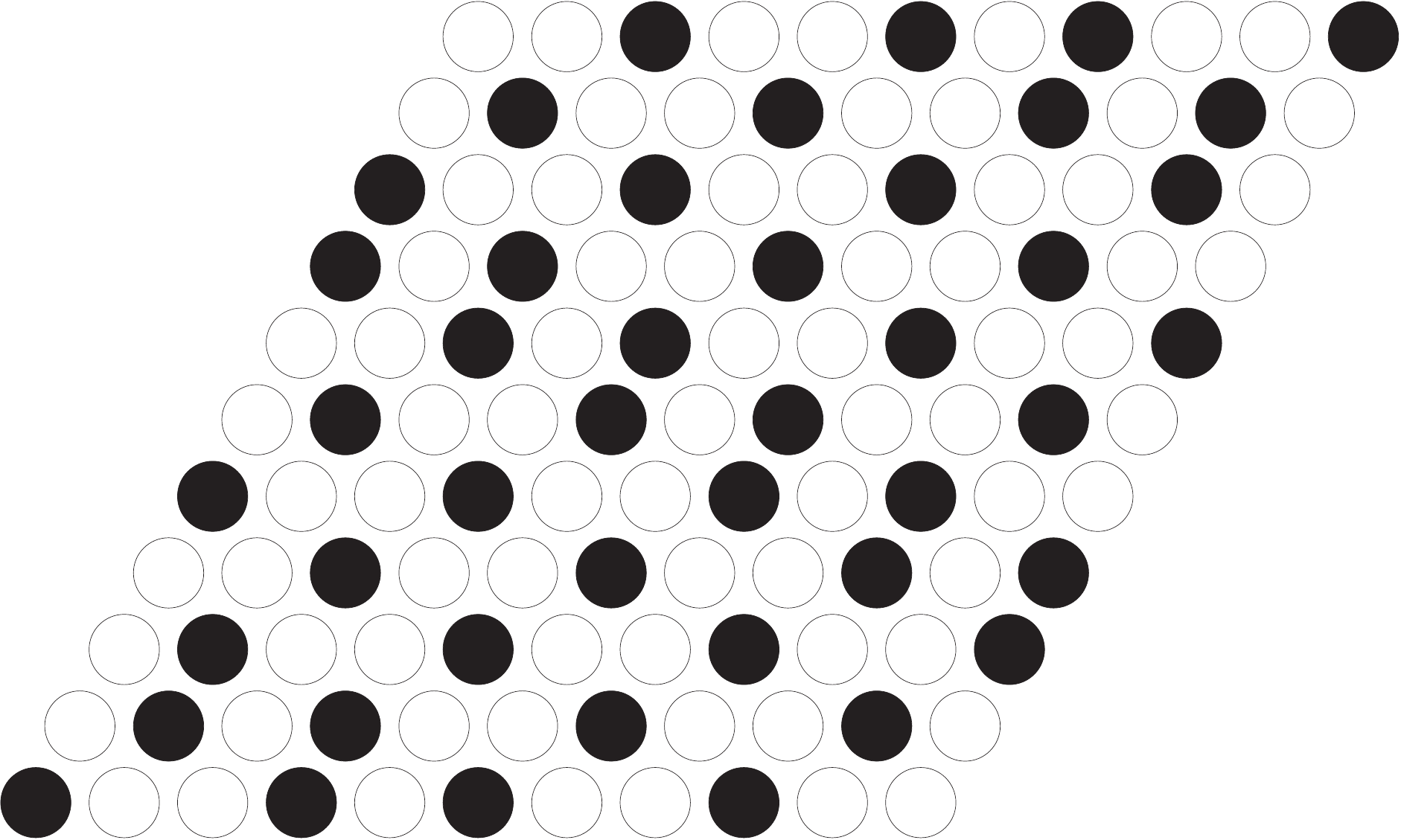}
}\hfill
\subfigure[\ $h = 4.0$ \label{f:D6z:inf_p3}]
{
\includegraphics*[width=3cm]{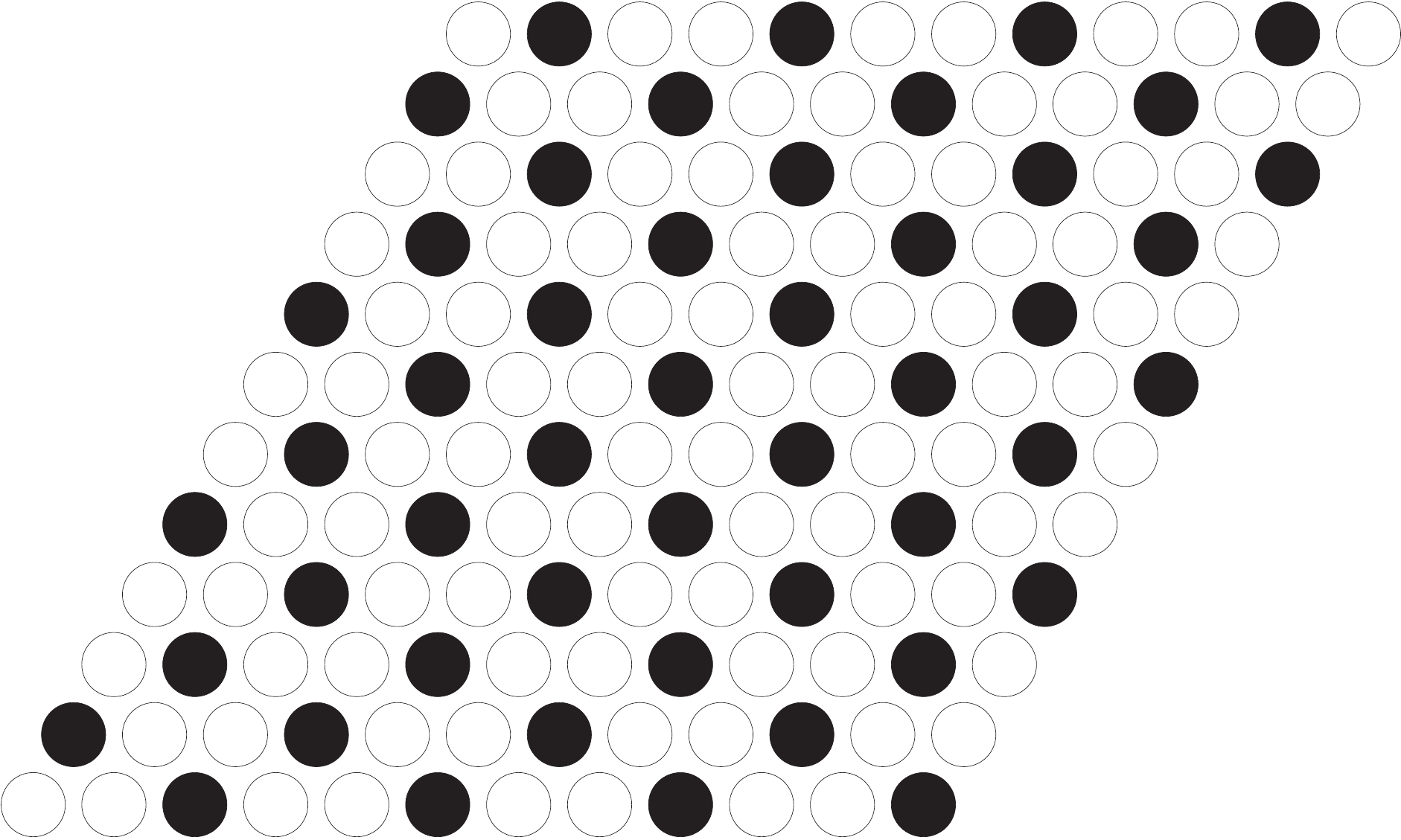}
}\hfill
\subfigure[\ $h = 6.77$ \label{f:D6z:inf_a3+}]
{
\includegraphics*[width=3cm]{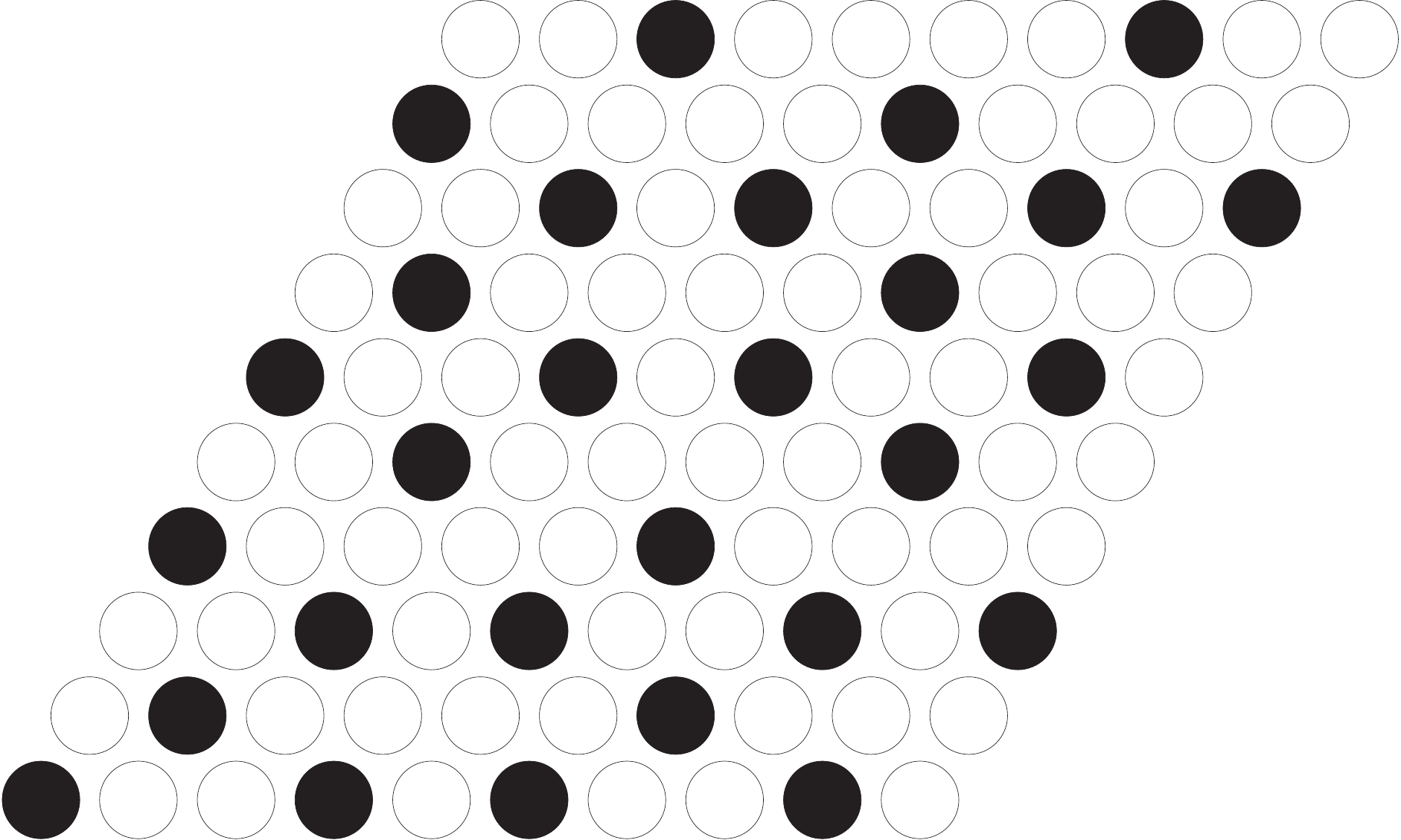}
}\hfill
\subfigure[\ $h = 7.5$ \label{f:D6z:inf_a4}]
{
\includegraphics*[width=3cm]{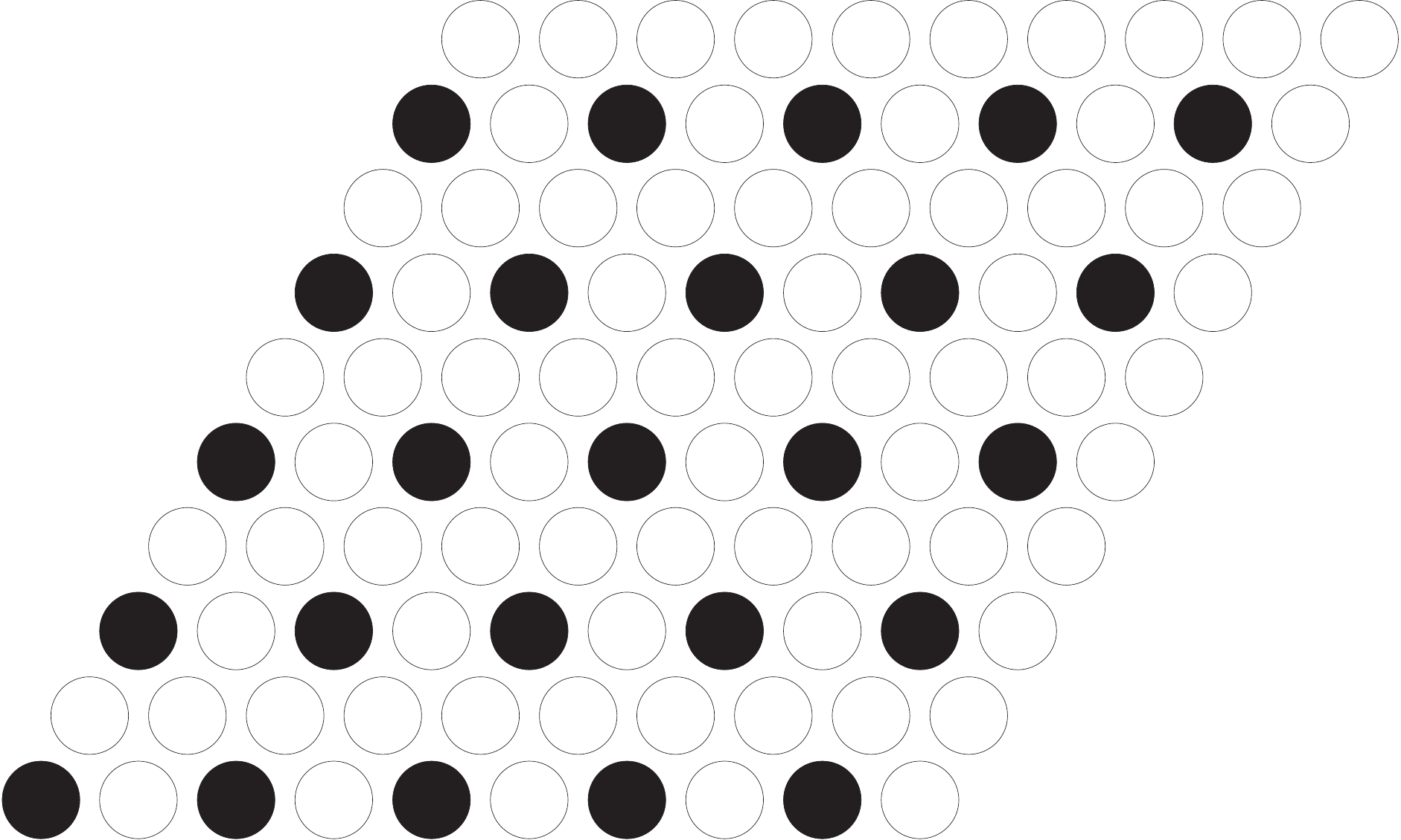}
}\hfill
\phantom{
\includegraphics*[width=3cm]{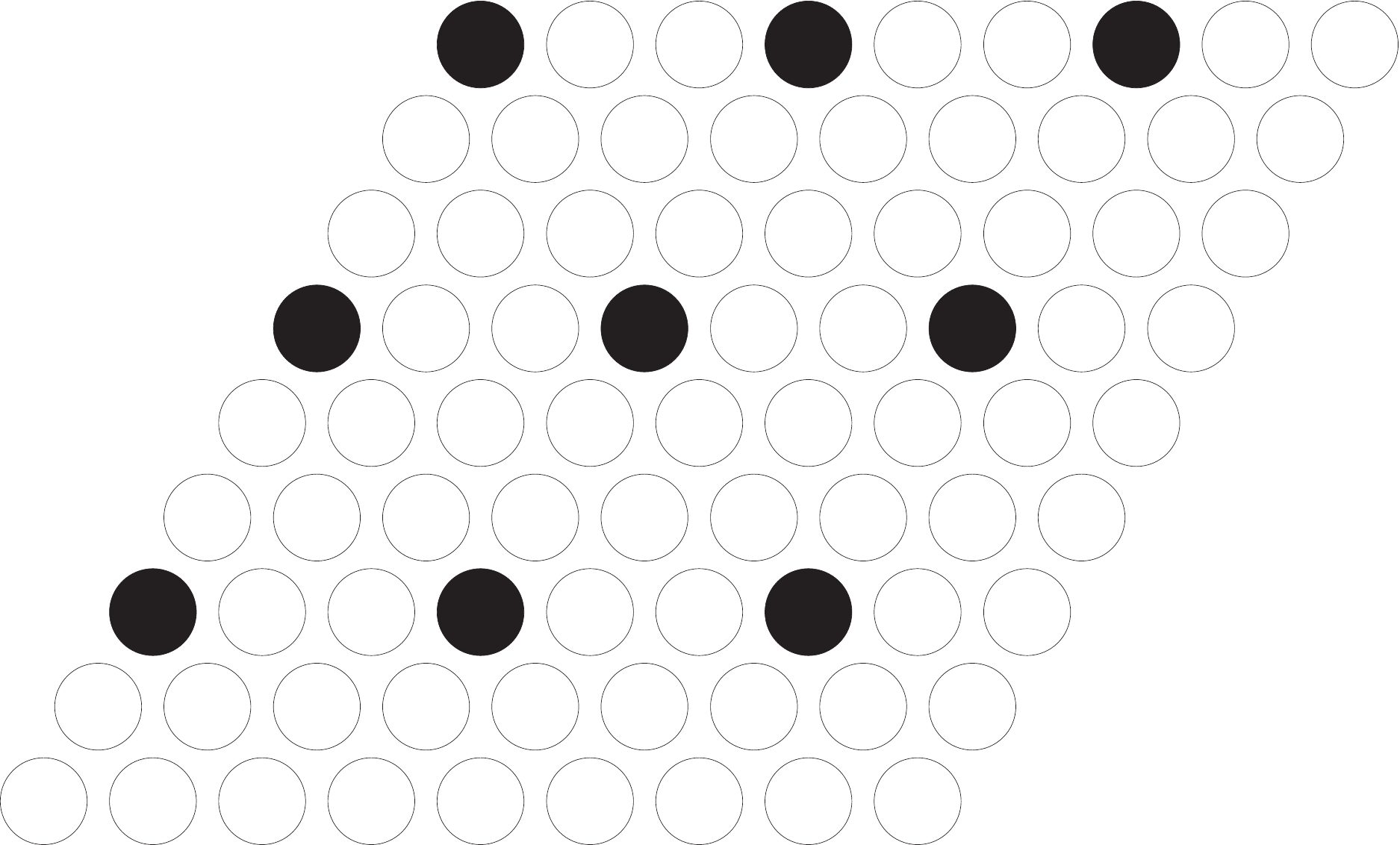}
}\hfill
\subfigure[\ $h = 9.5$ \label{f:D6z:inf_a7}]
{
\includegraphics*[width=3cm]{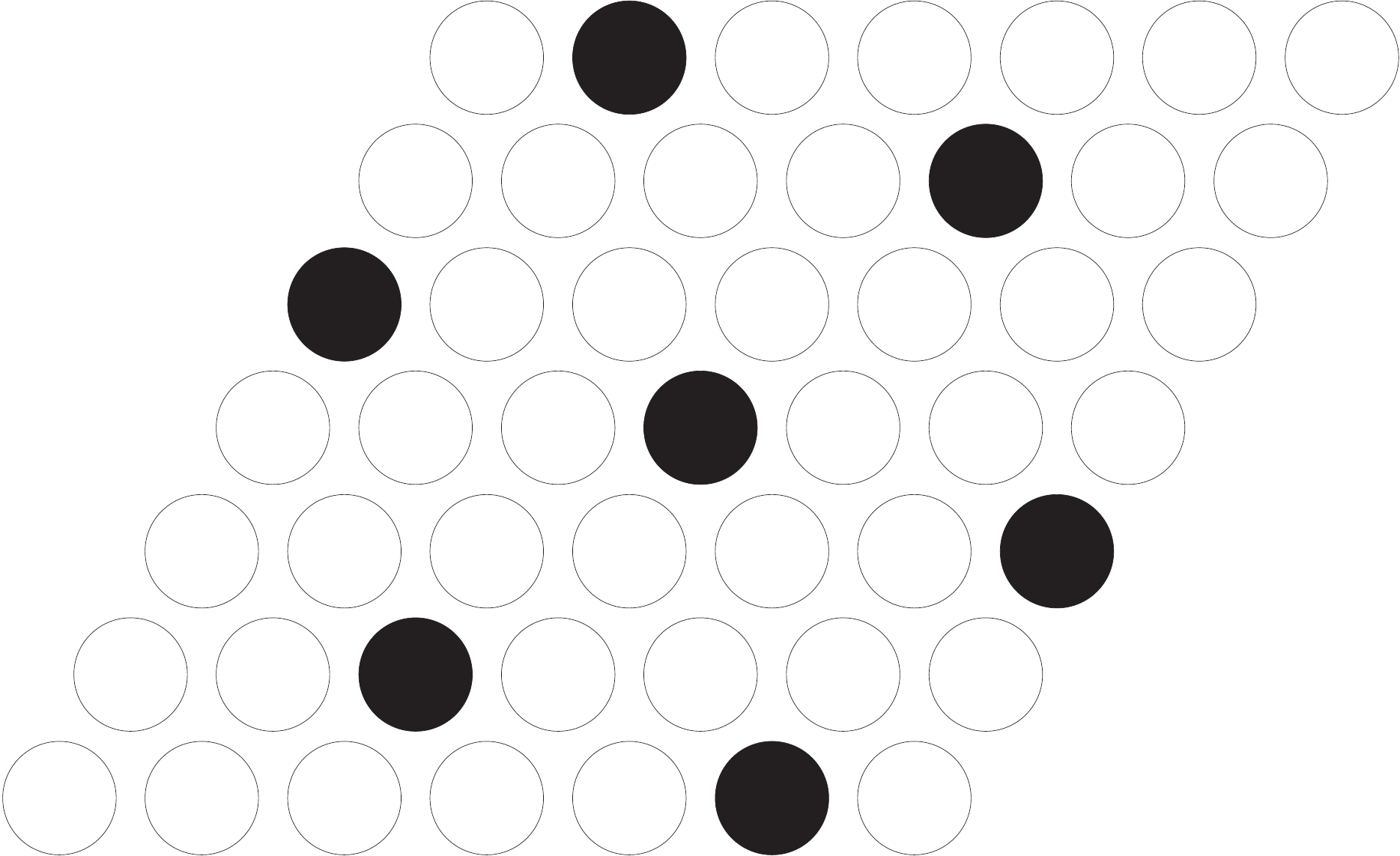}
}\hfill
\subfigure[\ $h = 10.1$ \label{f:D6z:inf_a9}]
{
\includegraphics*[width=3cm]{3j}
}\hfill
\phantom{
\includegraphics*[width=3cm]{3j}
}\hfill
\caption{Ground states for
characteristic values of magnetic field found by Monte-Carlo
simulations. \label{f:D6z:inf}}
\end{figure}

Note the specific regions of this dependence present at different
field intervals; first, the region with small values of $\langle m
\rangle \leqslant 0.2 m_0$ having a rather non-regular dependence of
$\langle m\rangle $ on $h$; second, the regions with constant values
of $\langle m \rangle$ independent of the magnetic field (shelves);
and third, the saturation region. The characteristic magnetic
structures found in these regions are depicted in figure~\ref{f:D6z:inf}.

Monte-Carlo data are not too clear in the region of small fields
such as $0.7\div0.9$, and the magnetic structures are far from the simple
AFM structures, see figures~\ref{f:D6z:inf_p9}--\ref{f:D6z:inf_p8}.
Within this region, the aforementioned structure with a set of
topological defects appears. A detailed discussion of this region is
the main issue of our article.

Within the shelf regions, almost all initial Monte-Carlo
configurations lead to the same magnetic structures, which
correspond to the formation of triangular superlattices for the
minority dots (antiparallel to the magnetic field) with different
lattice spacings. As an example, note the ideal  triangular
superlattices with  $\langle m \rangle = m_0/3$ and with the period
$a_\textrm{sl}/a = \sqrt{3}$, see figure~\ref{f:D6z:inf_p3} present at
the values $2.4 \lesssim h \lesssim 6.4$. For higher fields, the
superlattices with $a_\textrm{sl}/a = 2$ [figure~\ref{f:D6z:inf_a4}],
$a_\textrm{sl}/a = \sqrt{7}$, [figure~\ref{f:D6z:inf_a7}] and
$a_\textrm{sl}/a = 3$ [in figure~\ref{f:D6z:inf_a9}] correspond to such shelves.
For high magnetic fields near the saturation region, the
magnetization process is going through creating a superlattice of
flipped dots of small density.

In the region of low magnetic fields, as well as in the regions of
a magnetic field where the transitions between the superlattices
occur, resettability of Monte-Carlo result lowers, and the
results become unreliable. The observed magnetic structures in
these transition regions are characterized by much lower symmetry
than for the shelf regions.  For example, at the values $0.9
\lesssim h \lesssim 1.5$, where the finite (but small) magnetic
moment $\langle m_0 \rangle$ is formed, the translational symmetry
for a set of flipped dots cannot be attributed to a simple
superlattice structure, see figure~\ref{f:D6z:inf_p9}. However, in this
figure one can clearly see a novel element, an additional zigzag line of the sites
oriented parallel to one of the translation vectors of the lattice.
The resulting magnetic structure can be interpreted
as an antiferromagnetic domain structure in a system with a
zigzag line as a domain wall. Such topological linear defects were
described for a two-sublattice antiferromagnetic state with an
interaction of a few neighboring moments \cite{IvanovKireev09}. For
the region of small fields, an increase of magnetic field leads
to an increase of the density of topological linear defects, see
figures~\ref{f:D6z:inf_p9}--\ref{f:D6z:inf_p5}. A minimal field
for the start of this process corresponds to a low density of such
defects, and to find the critical field one needs to consider larger
and larger systems. Namely, to present a stripe of width $n$ we need
a system of at least $(2n + 1) \times (2n + 1)$ size. Below, in
section~\ref{sec:analytics} we find the starting field for the
creation of a set of topological defects by an analytical
calculation.

To refine the Monte-Carlo data and to clarify the
magnetic states at the fields of interest, $0 \leqslant h \leqslant 3$, we
perform a direct exhaustive search based on the picture of stripe
AFM domain structures for rhombus-shaped space regions with various
(not necessarily equidistant) geometries of domain lines, up to the
size $60 \times 60$. It appears that for all fields the only
equidistant structure corresponds to the minimal configurations,
with linear system of stripes at $h < 1.5$ or triangular
superlattice at $h > 2.0$. Then, the only equidistant structures with
the size up to $300 \times 300$ were examined. The magnetization
curve based on these calculations is represented above in the insert
in figure~\ref{f:D6z:magn_inf} and is compared with the Monte-Carlo
data.

The common ``topological'' scenarios are present for other
transition regions, both below and above the shelf regions with the
magnetic structure of a form of ideal  triangular superlattices of
minority dots. For example, the structure present at $2.0 \lesssim h
\lesssim 2.4$ can be described as a ``compression'' of the domains
of the superlattice of period $a\sqrt{3}$ by the lines of dots with
down magnetic moments, see figures~\ref{f:D6z:inf_p8}
and~\ref{f:D6z:inf_p11}, whereas the state at the opposite end of
this shelf can be seen as a ``rarefication'' of the $a\sqrt{3}$
superlattice, see figure~\ref{f:D6z:inf_a3+}. The transition
structures corresponding to the ``higher'' shelves have  $C_6$
symmetry, higher than for low field structures.

\section{Magnetic ground states: Analytic description}\label{sec:analytics}

For a non-frustrated square lattice of Ising magnetic moments, the
destruction of both types of states is started through the creation of
a point defect in the state, the single magnetic dot with the
magnetic moment being reversed with respect to the regular structure of a
given state \cite{BishGalkIv}. First, in this section we check
the validity of such a scenario for  saturated and AFM states of
a triangular lattice array of magnetic dots. Then, there follows a theoretical
description of the novel topological mechanisms.

\subsection{Point defect scenarios }

In order to determine the values of the magnetic fields which
correspond to  ``point defect instability'' we have calculated the
change in dipolar interaction energy that occurs when the magnetic
moment of a single dot is reversed with respect to the ferromagnetic
and two-sublattice AFM structures \cite{BishGalkIv}. This energy
change is determined by the energy per dot in the initial states,
which can be expressed by simple lattice sums calculated with high
precision. These sums here and below were calculated using a standard
software package \textit{Mathematica}.

The point defect scenario well describes the instability of the
saturated state. It is easy to see that the change of the energy of
the saturated state with the flip of a single magnetic moment can be
presented as $W_1 = 2m_0 (H -H_\textrm{sat})$,  where
\begin{equation}\label{D6z:h_sat}
H_\textrm{sat}= m_0\sum_{\mathbf{n}_{i} \neq 0}
\frac{1}{|\mathbf{n}_{i}|^{3}} \equiv h_\textrm{sat}\frac{m_0}{a^3} \,, \qquad
h_\textrm{sat} \approx 11.034176 \;.
\end{equation}

The quantity $H_\textrm{sat}$ determines the saturation field for a
triangular lattice of Ising magnetic moments. If the magnetic field
$H<H_\textrm{sat}$, the value of $W_1$ is negative and the flipping of
a dot becomes favorable.

In principle, the common calculations can be performed for AFM
state, as well as for any state with the superlattice of flipped
dots of the same symmetry as for underlying dot lattice. This
approach well describes the instability point for the AFM state for
a square lattice of magnetic dots \cite{BishGalkIv}. Its application
for the triangular dot lattice shows that the reversal of the
magnetic moment of one dot in the AFM state becomes favorable at
$H\geqslant H_{\mathrm{AFM}}$, with the value $h_{\mathrm{AFM}}
 = 1.8377$, is much higher than the instability
field $h \simeq0.7 $ found numerically in the previous section. Thus,
the reversal of a single magnetic moment cannot describe the
instability of the AFM state observed for a triangular lattice at $h
\leqslant 1$.

\subsection{Instability of AFM state through the creation of topological lines}
\begin{figure}[!t]
\begin{center}
\includegraphics[width=0.9\textwidth]{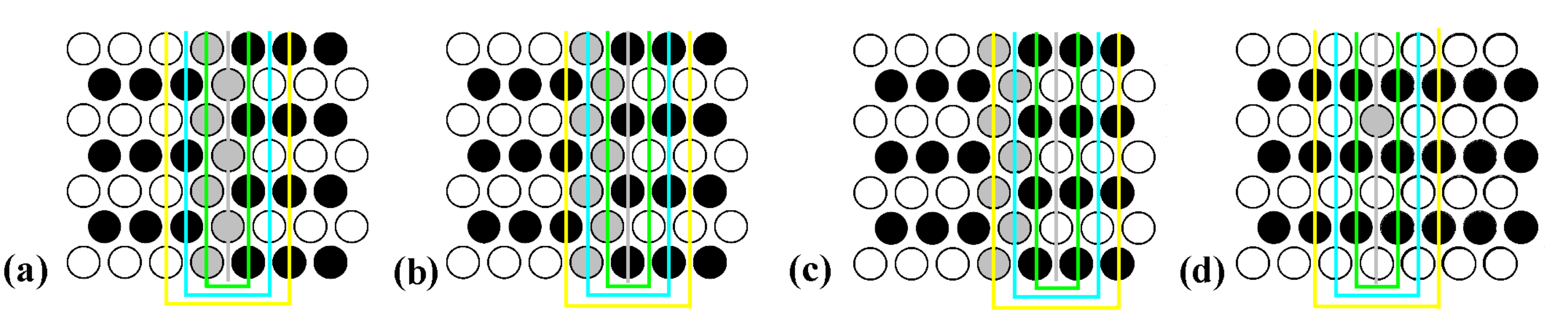}
\end{center}
\caption{(Color online). (a)--(c) The structure of topological
defect with the different sets of characteristic lines used for
calculation of the defect energy, see details in the text. (d) The
ideal AFM structure and the characteristic lines used for energy
calculation. As usually, the open and closed circles denote the
particles with the upward and downward magnetic moments,
respectively, but the dots with upward
magnetic moments within the defect line are mentioned by grey.} %
\label{line_dis} %
\end{figure}

As we found by Monte-Carlo analysis, the AFM state looses its
stability as the field increases due to the creation of
topological defect lines (domain walls). This  defect line
corresponds to an additional zigzag line of the particles with
magnetic moments, which are oriented as their neighbors, and should
be normal to one of the elementary translation vectors, see
figure~\ref{line_dis}.

The importance of defect lines in thermodynamics is a well-known
property of two-dimensional systems with a discrete symmetry breaking.
Due to the creation of a finite density of such lines, the long
range order is destroyed at finite temperature determined by the
energy of the defect, see, for example, \cite{KorshunovUFN}. However, for
frustrated AFM states, the behavior can be very unusual. In
particular, there is no magnetic order for AFM Ising system with the
nearest-neighbor interaction at any finite temperature $T
> 0$ \cite{NoPT,NoPT2,NoPT3,GekhtObz}. In general, this behavior can be explained
using the defect line picture of the phase transition, with the
vanishing of energy (more exactly, the free energy) of a certain
linear topological defect \cite{Korshunov}.

For our system of mesoscopic magnetic particles with a dipole
interaction, the  effect of a magnetic field, instead of thermal
effects, should be significant. The topological defect line with
non-zero magnetization and having zero energy  in the
nearest-neighbor approximation was recently found
\cite{IvanovKireev09}. This defect coincides with that observed in
our numerical simulations, compare figure~\ref{line_dis} and
figures~\ref{f:D6z:inf_p9}--\ref{f:D6z:inf_p8} above. A physical
consequence of a nonzero magnetic moment is that for such a defect,
an additional energy gain $m_0H$ per defect particle appears in the
magnetic field $H$. Then, the defect energy decreases as the field
increases and becomes zero at $H = H_{\mathrm{DW}} \equiv
E_{\mathrm{DW}}/m_0$. For $H \geqslant H_{\mathrm{DW}}$, the finite
density of such defects will be present in the ground state. This is
exactly the scenario observed in our numerical simulations at
magnetic field at the range $0.7\div1.5$.

In order to find the critical value of the field $H_{\mathrm{DW}}$,
let us calculate the energy of the domain wall $E_{\mathrm{DW}}$. It
is convenient to divide the full lattice into lines of dots parallel
to the defect line, as it is shown in figure~\ref{line_dis}. The
energy of the system with domain wall (per one dot in the defect
line) can be presented as a sum over these lines as follows:
\begin{equation}
E=E_{\mathrm{GS}}+E_{\mathrm{DW}}=-\frac{1}{2}\sum_{n}m_0
\sigma_{n}H_n\,, \label{el_gs}
\end{equation}
where $E_{\mathrm{GS}}$ is the energy of the ground state, the
integer $n$ describes the distance $a_n$ of the given line from the
defect line, $a_n=an/2$, $\sigma_{n}=\pm 1$ gives the sign of the
moment for the $n$-th line, and $H_n$ is the magnetic field created
on the dot in the $n$-th line by other dots in the system. To find
the field $H_n$, it is convenient to group all other dots to pairs of
lines equidistant from the $n$-th line, as it is shown for
$n=0,1,2$ in figure~\ref{line_dis}(a), (b) and (c), respectively.
Let us enumerate these pairs by an integer $k$ so that the distance
between the $n$-th line and one component of the $k$-th pair is equal to
$ak/2$, the pairs with $k=1,2,3$ are presented in
figure~\ref{line_dis}. Then, the energy of the magnetic state with a
domain wall can be presented by a double sum, over $n>0$ and $k>0$.

It is easy to see that for any finite $n$, the only pairs with
limited $k<n$ contribute to the energy of the state with the domain
wall. For example, for the lines directly entering the defect line
[$n=0$, see figure~\ref{line_dis}(a)], the contributions of two lines
composing any pair cancel each other. For this line, the non-zero
contribution to the energy is given by the dots from the same line,
and we denote this contribution as $\varepsilon _0$. Then, for the line
with $n=1$, only one pair gives non-zero contribution, see
figure~\ref{line_dis}(b), and the energy can be written as
$\varepsilon _0- 2\varepsilon _1$. Similarly, for $n=2$, the energy
is $\varepsilon _0- 2\varepsilon _1+ 2\varepsilon _2$, see
figure~\ref{line_dis}(c), and so on. Finally, the energy of the state
with a domain wall is presented through $\varepsilon _0$ and the
particular finite sums of the positive quantities $\varepsilon _n$,
\begin{equation}\label{Epsn} \nonumber
\varepsilon _{2n+1}=\sum_{k=1}^\infty \frac{4}{[(n+1/2)^2 +
3(k-1/2)^2]^{3/2}}\,, \qquad \varepsilon _{2n} = \sum_{k=1}^\infty
\frac{4}{(n^2 + 3k^2)^{3/2}}+\frac{2}{n^3}\,.
\end{equation}

The energy of the domain wall equals the difference of the energy of
the state with the domain wall, equation \eqref{el_gs} and the ground
state energy $E_{\mathrm{GS}}$. To find the ground state energy, it
is convenient to use the same presentation by a parallel lines,
see figure~\ref{line_dis}(d), and to present it by the same sums
$\varepsilon _n$. It is clear that the energy per one dot in any
line in the ground state is proportional to an infinite sum of the
form $\varepsilon _{\mathrm{GS}}=\varepsilon
_0+2\sum_{n=1}^{\infty}(-1)^n\varepsilon _n$. Then, the domain wall
energy can be found by term-by-term summation of the corresponding
contributions of the form $[(\varepsilon _{0}-\varepsilon
_{\mathrm{GS}})+(\varepsilon _{0}-2\varepsilon _{1}-\varepsilon
_{\mathrm{GS}})+\ldots]\equiv h_{\mathrm{DW}}=2\varepsilon
_{1}-4\varepsilon _{2}+6\varepsilon _{3}+\ldots$\,. The corresponding
infinite series $h_{\mathrm{DW}}=-2\sum_{n=1}^{\infty}(-1)^n n
\varepsilon _n$ are sign-alternating and converge quite well.
Finally, the domain wall energy per one dot $E_{\mathrm{DW}}$ can be
presented as follows:
\begin{equation}\label{Edw}
E_{\mathrm{DW}}=m_0 H_{\mathrm{DW}}\,, \qquad
H_{\mathrm{DW}}=h_{\mathrm{DW}}\frac{m_0}{a^3}\,, \qquad
h_{\mathrm{DW}}=0.70858944\,.
\end{equation}

Here, we also present the characteristic value of the magnetic field,
and $H_{\mathrm{DW}}=E_{\mathrm{DW}}/m_0$ determining the border of
stability of the simple AFM state; for $H>H_{\mathrm{DW}}$, AFM state
becomes unstable against the creation of domain walls. Note, that the
calculated value \eqref{Edw} is in good agreement with that found by
numerical simulations, but it is much lower than the field of point
defect instability for AFM state, $H_{\mathrm{AFM}}= 1.8377m_0/a^3$.

\subsection{Plateau description}

Monte-Carlo simulations show some peculiarities  (see
figure~\ref{f:D6z:magn_inf}) in the dependence of the magnetization on
the applied magnetic field in the form of plateaus, where the value of
the function does not change over a wide range of the argument.
These peculiarities have a simple explanation. The magnetization of
the array increases at a small external field due to the formation of
parallel topological defects in the form of domain walls. At some
critical concentration of such walls, the resulting state is nothing
but the superlattice of flipped dots which has a triangular
structure that coincides with the array symmetry; see
figure~\ref{f:D6z:inf}(f), (h), (i), (j). Such a superlattice transforms into
a self-similar structure but with another step (lattice
constant) as the applied field increases. Since the superlattice
constant has discrete values  $a_\textrm{sl}/a=\sqrt{3}$, 2, $\sqrt{7}$, 3, $2\sqrt{3}$, $\sqrt{13}$, $4$, $\sqrt{19}$, $\sqrt{21}$, $5\ldots$\,,
such a superstructure has good stability against the alteration
of the external field, and magnetization can be changed only
stepwise. One can see that such a structure consists of two
inversely magnetized  states with the lattice constants $a$ and
$a_\textrm{sl}$ and with the magnetization $\langle m\rangle
=m_0-2m_0(a/a_\textrm{sl})^2$. The value of the field of the stability
loss of such a superstructure relative to the transition to another
lattice constant can be easily calculated on the same principle as
the field of the  saturated state stability which was done above,
\begin{equation}\label{eqn_step}
H_\textrm{sl}=h_\textrm{sat}\frac{m_0}{a^3}
\left[1-2\left(\frac{a}{a_\textrm{sl}}\right)^3\right],
\end{equation}
where the multiplier $2$ in the numerator responds to the change of
magnetization of the dot in comparison with the saturated state and
$h_\textrm{sat}=11.034176$ is the field of the transition to the
saturated state. Though this effect occurs in a narrow range of
the field, it leads to instability of the superlattice at a little
bit smaller value of the field than it is predicted in equation
(\ref{eqn_step}).

\section{Conclusions}

The triangular regular array of magnetic particles demonstrates
quite non-trivial scenario of phase transitions in the presence
of an external magnetic field. The combination of two kinds of the
origin of frustration, the first, present for non-bipartite triangular
lattice with any kind antiferromagnetic interaction, and the second,
connected with the long-range character of magnetic dipole
interaction, leads to the creation of a set of linear topological
defects with the growth of the magnetic field.
Self-organization of such defects determines the magnetization
process for a system within a wide range of external
magnetic fields. It is worth noting an essential difference from
the standard topological transitions known for an atomic spin system
of low-dimensional magnets. For a system of microparticles (atomic
spins) of standard low-dimensional magnets, the topological defects
(vortices, domain walls, etc.) appear as a consequence of thermal
fluctuations, the resulting structure being quite irregular. By
contrast, in our case of the system of mesoscopic elements, the
characteristic energy of the interaction of two particles is much
higher than the Curie temperature. The creation/annihilation of the
defects takes place at the critical value of the magnetic field,
where the energy of the defect vanishes, whereas other characteristics
(such as the typical interaction energy of the defects) are still finite
and high, up to the values of 10$^5$ Kelvin \cite{BishGalkIv}. In our
case, a quite regular pattern of these defects is realized in the
system even at finite temperatures below the Curie temperature.

\vspace{-3mm}

\section*{Acknowledgements}
{This work was partly supported by the State Foundation of
Fundamental Research of Ukraine via grant No.~F53.2/045}

\vspace{-7mm}

\ukrainianpart

\title{Самоорганізація топологічних дефектів в трикутній гратці магнітних точок під впливом перпендикулярного магнітного поля}
\author{Р.С. Химин\refaddr{label2}, В.Є. Кірєєв\refaddr{label1},
Б.О. Іванов\refaddr{label1}}
\addresses{
\addr{label1} Інститут магнетизму НАН України та МОН України, бульв.
Вернадського, 36-б,  03142 Київ, Україна
\addr{label2} Університет Окленд, Рочестер Хіллс, 48309 Мічиган, США}
%
%
%

 \makeukrtitle

\begin{abstract}
\tolerance=3000

Періодичний масив магнітних частинок (магнітних точок), сформований
в трикутну двовимірну гратку, утворює складну струтуру магнітного
впорядкування. Магнітна симетрія основного стану для такої системи
виявляється нижчою ніж симетрія гратки. Далекодіюча диполь-дипольна
взаємодія призводить до специфічного антиферомагнітного
впорядкування при малих полях, в той час як при збільшенні
магнітного поля виникає серія лінійних топологічних дефектів.
Самоорганізація таких дефектів визначає процес намагнічування
системи в широкому діапазоні зовнішнього магнітного поля.

\keywords магнітна точка, топологічний дефект

\end{abstract}

\end{document}